\documentclass[%
 reprint,
 amsmath,amssymb,
 aps,
]{revtex4-1}

\usepackage{graphicx}
\usepackage{dcolumn}
\usepackage{bm}
\usepackage{color}

\include{epsf} 
\newtheorem{Def}{Definition}
\newtheorem{Remark}[Def]{Remark}
\newtheorem{Ex}[Def]{Example}

\newcommand{\N}{{\rm I\! N}}
\newcommand{\R}{{\rm I\! R}}

\newcommand{\QED}{\hfill $\Box$}

\begin{document}


\title{A semi-Markov model for price returns}
\author{Guglielmo D'Amico}
\affiliation{Dipartimento di Scienze del Farmaco, Universit\'a G. D'Annunzio, Chieti}
\author{Filippo Petroni}
\affiliation{Dipartimento di Scienze Economiche e Commerciali, Universit\'a di Cagliari}%
 

\date{\today}

\begin{abstract}
We study the high frequency price dynamics of traded stocks by a model of returns using a semi-Markov approach.
More precisely we assume that the intraday return are described by a discrete time homogeneous semi-Markov process and the overnight returns are modeled by a Markov chain.  Based on this assumptions we derived the equations for the first passage time distribution and the volatility autocorreletion function. Theoretical results have been compared with empirical findings from real data. In particular we analyzed high frequency data from the Italian stock market from first of January 2007 until end of December 2010.  The semi-Markov hypothesis is also tested through a nonparametric test of hypothesis.
\end{abstract}

\pacs{Valid PACS appear here}
\maketitle


\section{Introduction}
Semi-Markov processes (SMP) are a wide class of stochastic processes which generalize at the same time both Markov chains and renewal processes. Their main advantage is that of using whatever type of waiting time distribution for modeling the time to have a transition from one state to another one. This major flexibility has a price to pay: availability of data to estimate the parameters of the model which are more numerous. Semi-Markov processes generalizes also non-Markovian models based on continuous time random walks used extensively in the econophysics community, see for example \cite{mai00,rab02}.   
SMP have been used to analyze financial data and to describe different problems ranging from credit rating data modeling \cite{dam05} to the pricing of options \cite{dam09,sil04}.

With the financial industry becoming fully computerized, the amount of recorded data, from daily close all the way down to tick-by-tick level, has exploded. Nowadays, such tick-by-tick high-frequency data are readily available for practitioners and researchers alike\cite{gui97,pet03}. It seemed then natural to us trying to verify the semi-Markov hypothesis of returns on high-frequency data.
 
We propose a semi-Markov model for price return. More precisely we assume that the intraday returns (up to one minute frequency) are described by a discrete time homogeneous semi-Markov process and the overnight returns are modeled by a Markov chain. In this way we can consider differently the intraday and the overnight activities. 
To establish the validity of our model we tested it first of all by using a nonparametric test proposed by \cite{ste06} and then against two of the stylized facts which characterized financial data: the first passage time distribution  \cite{sim02,jen04} and the autocorrelation function of the square of returns.

Following the model we determine equations for the first passage distributions and the intraday autocorrelation function by using renewal type arguments.
Results from the model are then compared with empirical results obtained from the data.
We show that these stylized facts are better reproduce when the semi-Markov model is used compare to a simple Markov chain.
 
The database used for the analysis is made of high frequency tick-by-tick price data from all the stock in Italian stock market from first of January 2007 until end of December 2010. From prices we then define returns at one minute frequency.

The paper is divided as follows: First, semi-Markov processes, notation and some results are described in Section 2. Next, the price model is illustrated and first passage time distributions and the intraday autocorrelation functions are computed in Section 3. Finally, in Section 4, an application to real high frequency data illustrates the results.

\section{Semi-Markov Processes}
We define an HSMP with values in a finite state space $E=\{1,2,...,m\}$, see for example \cite{lim01,jan06}. Let $(\Omega,\mathbf{F},P)$ be a probability space; we consider two sequences of random variables:
\begin{equation*}
J_{n}:\Omega\rightarrow E\,;\,\,\,\,\,\,T_{n}:\Omega \rightarrow \N
\end{equation*}
\noindent denoting, respectively, the state and the time of the n-th transition of the system.\\
\indent We assume that $(J_{n},T_{n})$ is a Markov Renewal Process on the state space $E \times \N$ with kernel $Q_{ij}(t),\,\,i,j\in E, t\in \N$.\\
\indent The kernel has the following probabilistic interpretation:
\begin{equation}
\label{due}
\begin{aligned}
&P[J_{n+1}=j, T_{n+1}-T_{n}\leq t |\sigma(J_{h},T_{h}),\,h\leq t, J_{n}=i]=\\
& P[J_{n+1}=j, T_{n+1}-T_{n}\leq t |J_{n}=i]=Q_{ij}(t)
\end{aligned}
\end{equation}
and it results $p_{ij} = \mathop {\lim }\limits_{t\, \to \,\infty } Q_{ij}(t); \, i, j \in E, \, t\in \N$
where
$
{\bf P} = (p_{ij})
$
is the transition probability matrix of the embedded Markov chain $J_{n}$.\\
\indent  Furthermore, it is useful to introduce the probability to have next transition in state $j$ at time $t$ given the starting at time zero from state $i$
\begin{eqnarray}
\label{tre}
&&b_{ij}(t)=P[J_{n+1}=j, T_{n+1}-T_{n}= t |J_{n}=i]= \nonumber \\
&&=\left\{
                \begin{array}{cl}
                       \ Q_{ij}(t)-Q_{ij}(t-1)  &\mbox{if $t>0$}\\
                         0  &\mbox{if $t=0$}\\
                   \end{array}
             \right.
\end{eqnarray}
\indent We define the distribution functions
\begin{equation}
\label{trebis}
H_{i}(t)=P[T_{n+1}-T_{n}\leq t |J_{n}=i]=\sum_{j\in E}Q_{ij}(t)
\end{equation}
\noindent representing the survival function in state $i$.\\
\indent The Radon-Nikodym theorem assures for the existence of a function $G_{ij}(t)$ such that
\begin{eqnarray}
\label{quattro}
&&G_{ij}(t)=P\{T_{n+1}-T_{n}\leq t|J_{n}=i, J_{n+1}=j\}=\nonumber \\ 
&&\left\{
                \begin{array}{cl}
                       \ \frac{Q_{ij}(t)}{p_{ij}}  &\mbox{if $p_{ij}\neq 0$}\\
                         1  &\mbox{if $p_{ij}=0$}\\
                   \end{array}
             \right.
\end{eqnarray}
\indent It denotes the waiting time distribution function in state $i$ given that, with next transition, the process will be in the state $j$. The sojourn time distribution $G_{ij}(\cdot)$ can be any distribution function. We recover the discrete time Markov chain when the $G_{ij}(\cdot)$ are all geometrically  distributed.\\
\indent It is possible to define the HSMP $Z(t)$ as
\begin{equation}
\label{cinque}
Z(t)=J_{N(t)},\,\,\,\forall t\in \N
\end{equation}
\noindent where $N(t)=\sup\{n\in \N: T_{n}\leq t\}$. Then $Z(t)$ represents the state of the system for each waiting time. We denote the transition probabilities of the HSMP by $\phi_{ij}(t)=P[Z(t)=j|Z(0)=i]$. They satisfy the following evolution equation:
\begin{equation}
\label{sei}
\phi_{ij}(t)=\delta_{ij}(1-H_{i}(t))+\sum_{k\in E}\sum_{\tau =1}^{t}b_{ik}(\tau)\phi_{kj}(t-\tau).
\end{equation}
\indent To solve equation $(\ref{sei})$ there are well known algorithms in the SMP literature \cite{bar04, jan06}.\\
\indent At this point we introduce the discrete backward recurrence time process linked to the SMP. For each time $t\in \N$ we define the following stochastic process:
\begin{equation}
\label{diciannove}
   B(t)=t-T_{N(t)}.
\end{equation}
\indent We call it discrete backward recurrence time process.\\
\indent If the semi-Markov process $Z(t)$ indicates the state of the system at time $t$, $B(t)$ indicates the time since the last jump.
\begin{figure}[h]
\label{traiettoria}
\centering
\includegraphics[width=8cm]{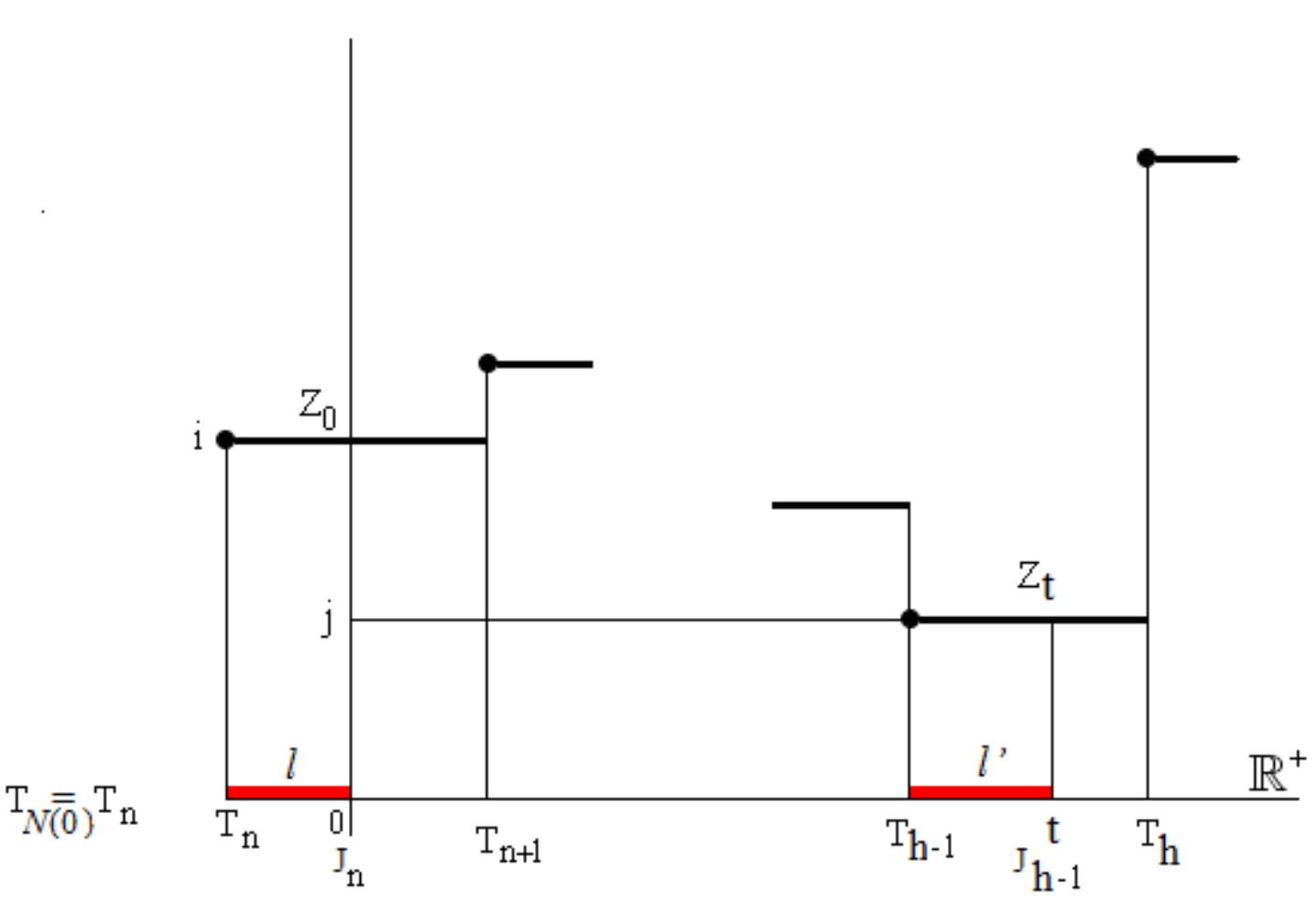}
\caption{Trajectory of a HSMP with backward times}
\end{figure}

\indent In Figure 1 we show a trajectory of an HSMP. At time $t$ the process $Z(t)$ is in the state $J_{h-1}$ and the last transition occurred at time $T_{h-1}$ then at time $t$ the backward process holds $B(t)=t-T_{h-1}$.\\
The joint stochastic process $(Z(t),B(t), t\in \N)$ with values in $E\times \N $ is a Markov process, see for example \cite{lim01}. That is:
\begin{displaymath}
\begin{aligned}
& P[Z(T)\!=\!j, B(T)\! \leq \!v'|\sigma(Z(h),B(h)),h\! \leq \!t, Z(t)\!=\!i, B(t)\!=\!v]\\
& =P[Z(T)=j, B(T)\leq v'|Z(t)=i, B(t)=v].
\end{aligned}
\end{displaymath}
To safe space let denote the event $\{Z(0)=i, B(0)=v\}$ in a more compact form by $(i,v)$.\\
\indent In the sequel of the paper we will make use of the following probabilities:
\begin{equation}
\begin{aligned}
& ^{b}\phi_{ij}^{b}(v;v',t)=P[Z(t)=j,B(t)=v'|(i,v)];\\
& ^{b}\phi_{ij}(v;t)=P[Z(t)=j|(i,v)].
\end{aligned}
\end{equation}

\indent Our next step is to compute $^{b}\phi_{ij}^{b}(v;v',t)$ as a function of the semi-Markov kernel. The results here below have been proved in \cite{dam09} and further developed in \cite{dam10}.\\
\indent For all states $i,j\in E$ and times $h,v,t \in \N$ such that $H_{i}(v)<1$ we have that\\
\begin{equation}
\label{venti}
\begin{aligned}
 ^{b}\phi_{ij}^{b}(v;v',t)=&\frac{\delta_{ij}[1-H_{i}(t+v)]}{[1-H_{i}(v)]}1_{\{v'=t+v\}}+\\
& \sum_{k\in E}\sum_{s=1}^{t}\frac{b_{ik}(s+v)}{[1-H_{i}(v)]}\,\,^{b}\phi_{kj}^{b}(0;v',t-s),
\end{aligned}
\end{equation}

Notice that $^{b}\phi_{ij}^{b}(0;v',t)$ satisfy the following system of equations
\[
^{b}\phi_{ij}^{b}(0;v',t)\!\!=\!\!\delta_{ij}[1\!-\!H_{i}(t+v)]\!+\!\sum_{k\in E}\!\sum_{s=1}^{t}\!b_{ik}(s+v)\,^{b}\phi_{kj}^{b}(0;v'\!,t\!-\!s).
\]
\indent This system can be solved with algorithms similar to that used for equation $(\ref{sei})$.

It results that $^{b}\phi_{ij}(v;t)=\sum_{v'=0}^{t+v} {^{b}\phi_{ij}^{b}(v;v',t)}$. Consequently:
\begin{equation}
\label{ventuno}
\begin{aligned}
& ^{b}\phi_{ij}(v;t)=\delta_{ij}\frac{[1-H_{i}(t+v)]}{[1-H_{i}(v)]}\\
& +\sum_{k\in E}\sum_{s=1}^{t}\frac{b_{ik}(s+v)}{[1-H_{i}(v)]}\phi_{kj}(t-s)
\end{aligned}
\end{equation}

\section{The price model}\label{price}

Let us assume that the value of the asset under study is described by the time varying asset price $S(t)$. The time variable $t\in \{0, 1, \ldots, nd \}$ where $n$ is the number of unit periods during the day (i.e. minutes) and $d$ is the number of days.
 
The intraday return at time $t$ calculated over a time interval of length 1, is defined as
\begin{equation}
Z(t) = \frac{S(t + 1) - S(t)}{S(t)}
\end{equation}
while if $t=nk$, $k=1,2,...,d$ we define the overnight return as
\begin{equation}
X(t) = \frac{S(t + 1) - S(t)}{S(t)}.
\end{equation}
\indent We assume that $Z(t)$ is a discrete time HSMP with finite state space 
$$E=\{-z_{min}\Delta,\ldots,-2\Delta,-\Delta,0,\Delta,2\Delta,\ldots, z_{max}\Delta\}$$
and kernel ${\bf{b}}=(b_{ij}(\gamma))$, $\forall i,j\in E$ and $\gamma \in \N $. On the contrary we describe $X(t)$ as a discrete time homogeneous Markov chain with the same state space and transition probability matrix ${\bf{T}}=(t_{i,j})_{i,j\in E}$.

We made this choice to take into consideration two different types of market activity: one (intraday) when the market is open and the second one (overnight) when, even if the market is closed, the opening price reflects the information accumulated during the stop of the activity.    
We define a simplified expression taking into account both the intraday and overnight returns: 
\begin{equation}
W(t)=\left\{
                \begin{array}{cl}
                       \ Z(t)  &\mbox{if $(k-1)n<t<nk$}\\
                         X(t)  &\mbox{if $t=nk$}\\
                   \end{array}
             \right.
\end{equation}
where $k=1,2,...,d$.

\subsection{The first passage time distributions}

The accumulation factor from $t$ to $t+\tau$ is given by
\begin{equation}
M_{t}(\tau) = \prod_{k=0}^{\tau-1}\left(1+W(t+k)\right).
\end{equation}
\noindent and takes value in the set $$SP_{\tau}=\{x\in \R : x=\prod_{k=0}^{\tau-1}\left(1+i(t+k)\right), i(t+k)\in E\}.$$
\indent More in general we need to introduce the symbol $SP_{\tau}^{\rho}=SP_{\tau}\bigcap (-\infty, \rho)$ to denote the set of accumulation factor values being less than $\rho$ at time $\tau$.\\ 
\indent It is easy to verify the relation between $M_{t}(\tau)$ and the price $S(t)$ 
\begin{equation}
M_{t}(\tau)=\frac{S(t + \tau)}{S(t)}.
\end{equation}

The fpt for an investment made at time $t$ at price $S(t)$, is defined as the time interval $\tau = t'-t$, $t'>t$, where the relation $M_{t}(\tau) \geq \rho$ is fulfilled for the first time. We will denote the fpt as $\lambda_\rho(t)$. Then $$\lambda_\rho(t)=\min\{\tau \geq 0 ; M_{t}(\tau) \geq \rho\}.$$

We assume that the semi-Markov process $Z(t)$ is time homogeneous then we can simply denote the fpt $\lambda_\rho(t)=\lambda_\rho$. We are interested in finding the distributional properties of the fpt. For each time $t$, let $$R_{i}(v,t;\rho)=P(\lambda_\rho >t|(i,v))$$
where $i\in E$ denotes the state of the return and $v\in \N$ the time length of being in this state both at time zero.

Let us define by $R_{i,j}(v,t;w,\rho)$, $\forall w\in SP_t$, $\forall \rho \in \R$, the probability 
$$
P(\lambda_\rho > t, W(t)=j, M_{0}(t+1)=w|(i,v)),     
$$
obviously 
\begin{equation}
\label{numerouno}
R_{i}(v,t;\rho)=\sum_{j\in E}\sum_{x\in SP_{t}, x<\rho}R_{i,j}(v,t;x,\rho).
\end{equation}
Here below we derive the equation for finding the fpt distribution in the proposed model. In the following we should distinguish different cases.
The first case if for time $1\leq t \leq n-1$. This means that we are interested in determining $R_{i,j}(v,T;w,\rho)$ for time $t$ belonging to the first day.\\
Being the events $\{T_{1}^{w}=k\}$ disjoint, it follows that 
\begin{equation}
\label{uno}
\begin{aligned}
& R_{i,j}(v,t;w,\rho)\\
& =P(\lambda_\rho > t, W(t)=j, M_{0}(t+1)=w, T_{1}^{w}>t|(i,v))\\
& +P(\lambda_\rho > t, W(t)=j, M_{0}(t+1)=w, T_{1}^{w}\leq t|(i,v)).
\end{aligned}
\end{equation} 
First addend on the right hand side (r.h.s.) of (\ref{uno}) is equal to:
\begin{equation*}
\begin{aligned}
& P(\forall u \in (0,t+1], M_{0}(u)<\rho| W(t)=j, T_{1}^{w}>t,(i,v))\\
& \cdot P(W(t)=j| T_{1}^{w}>t,(i,v))\cdot P(T_{1}^{w}>t|(i,v))\\
& =1_{\{(1+i\Delta)^{t}<\rho\}}\delta_{ij}\bigg(\frac{1-H_{i}(t+v)}{1-H_{i}(v)}\bigg).
\end{aligned}
\end{equation*}
Second addend on the r.h.s. of (\ref{uno}) is equal to:
\begin{equation*}
\begin{aligned}
& \sum_{a\in E}\sum_{m=1}^{t}P(\forall u\in (0,t+1], M_{0}(u)<\rho,\\
& M_{0}(t+1)=\!w, W(t)\!=\!j,T_{1}^{w}\!=\!m, J_{1}^{w}\!=\!a|(i,v))\\ 
& =\sum_{a\in E}\sum_{m=1}^{t}\!P(\forall u\! \in \!(m,t\!+\!1], M_{0}(m)M_{m}(u)< \rho,\\
& M_{0}(m)M_{m}(t+1)=w, W(t)=j|\forall u\in (0,m], \\
& M_{0}(u)<\rho, T_{1}^{w}=m, J_{1}^{w}=a)P(T_{1}^{w}\!=\!m, J_{1}^{w}\!=\!a|(i,v))\\
& \cdot P(\forall u\in (0,m], M_{0}(u)<\rho | T_{1}^{w}\!=\!m, J_{1}^{w}\!=\!a)\\ 
& =\sum_{a\in E}\sum_{m=1}^{t}\frac{b_{ia}(v+m)}{1-H_{i}(v)}1_{\{(1+i\Delta)^{m}<\rho\}} \\
& \cdot P(\forall u\in (m,t+1], (1+i\Delta)^{m}M_{0}(u-m)<\rho,\\
& (1+i\Delta)^{m}M_{0}(t+1-m)=w, W(t)\!=\!j|T_{1}^{w}\!=\!m, J_{1}^{w}\!=\!a)\\  
& =\sum_{a\in E}\sum_{m=1}^{t}\frac{b_{ia}(v+m)}{1-H_{i}(v)}1_{\{(1+i\Delta)^{m}<\rho\}} \\
& \cdot R_{a,j}\bigg(0,t-m;\frac{w}{(1+i\Delta)^{m}},\frac{\rho}{(1+i\Delta)^{m}}\bigg). 
\end{aligned}
\end{equation*}
This proves the following renewal-type equation for the fpt when horizon time $t$ belongs to the first day:
\begin{equation}
\label{fptI}
\begin{aligned}
& R_{i,j}(v,t;w,\rho)=1_{\{(1+i\Delta)^{t}<\rho\}}\delta_{ij}\bigg(\frac{1-H_{i}(t+v)}{1-H_{i}(v)}\bigg)+\\
& \sum_{a\in E}\sum_{m=1}^{t}\frac{b_{ia}(v+m)}{1-H_{i}(v)}1_{\{(1+i\Delta)^{m}<\rho\}}\cdot \\
& R_{a,j}\bigg(0,t-m;\frac{w}{(1+i\Delta)^{m}},\frac{\rho}{(1+i\Delta)^{m}}\bigg). 
\end{aligned}
\end{equation}
Now let us consider the case in which $t=n$. This means that we work until the opening of the second day. In this situation we should take care for the transition at time $t=n$ which is due to the Markov chain $X(t)$. To obtain a formula for the fpt until time $n$ it is sufficient to consider all possible states for the return and for the accumulation factor at time $n-1$ and to use equation (\ref{fptI}).
\begin{equation*}
\begin{aligned}
& R_{i,j}(v,n;w,\rho)=\\
&\sum_{\overline{w}\in SP_{n-1}^{\rho}}\sum_{a\in E}P(\forall u\in (0,n+1], M_{0}(u)<\rho,\\
& M_{0}(n+1)=\!w, W(n)\!=\!j, W(n-1)=a,\\
& M_{0}(n)=\!\overline{w}|(i,v))=\\
& \sum_{\overline{w}\in SP_{n-1}^{\rho}}\sum_{a\in E}P(M_{0}(n)(1+W(n))\!=\!w, W(n)\!=\!j| \\
& \forall u\in (0,n], M_{0}(u)<\rho, W(n-1)=a, M_{0}(n)=\!\overline{w},(i,v))\cdot \\
& P(\forall u\in (0,n], M_{0}(u)<\rho, W(n-1)=a, \\
& M_{0}(n)=\!\overline{w}|(i,v))=\\
& \sum_{\overline{w}\in SP_{n-1}^{\rho}}\sum_{a\in E} P((1+W(n))\!=\!\frac{w}{\overline{w}}, W(n)\!=\!j| W(n-1)=a) \cdot \\
& P(\forall u\in (0,n], M_{0}(u)<\rho, W(n-1)=a,M_{0}(n)=\!\overline{w}|(i,v)
\end{aligned}
\end{equation*}
\begin{equation}
\label{fptcaso2}
= \sum_{\overline{w}\in SP_{n-1}^{\rho}}\sum_{a\in E} t_{a,j}1_{\{(1+j\Delta)=\frac{w}{\overline{w}}\}} R_{i,a}(v,n-1;\overline{w},\rho)
\end{equation}
By similar computations it is possible to have the fpt distribution for time $t=nd$. The relation is the following:
\begin{equation}
\label{fptnk}
\begin{aligned}
& R_{i,j}(v,nd;w,\rho)=\\
& \sum_{a\in E}\sum_{\overline{w}\in SP_{(n-1)d}}^{\rho}\!\!\!\!\!\!\!\!R_{i,a}(v,(n-1)d;\overline{w},\rho)R_{a,j}(0,n;\frac{w}{\overline{w}},\frac{\rho}{\overline{w}}).
\end{aligned}
\end{equation}
Formula (\ref{fptnk}) is obtained by conditioning on all possible states of the return process $W(t)$ and on all possible values of the accumulation factor $M_{0}(t)$ at time $t=(n-1)d$ (the closing of day $n-1$).\\
The last case, when $(n-1)d<t<nd$, can be obtained by using jointly formulae (\ref{fptI}) and (\ref{fptnk}). The resulting relation is the following:
\begin{equation}
\label{fptnk-}
\begin{aligned}
& R_{i,j}(v,t;w,\rho)=\\
& \sum_{a\in E}\sum_{\overline{w}\in SP_{(n-1)d}}^{\rho}\!\!\!\!\!\!\!\!\!R_{i,a}(v,(n\!-\!1)d;\overline{w},\rho)R_{a,j}(0,t\!-\!(n\!-\!1)d;\frac{w}{\overline{w}},\frac{\rho}{\overline{w}}).
\end{aligned}
\end{equation}
Formula (\ref{fptnk-}) is obtained by conditioning on the states of the return process and on the values of the accumulation factor process at time $t=(n-1)d$ and then by using formula (\ref{fptI}).\\
\indent Formulae (\ref{fptI}), (\ref{fptcaso2}), (\ref{fptnk}) and (\ref{fptnk-}) allow us to compute the probability $R_{i,j}(v,t;w,\rho)$ for all times $t$. It should be noted that if $\rho$ is not too much big, it is highly probable that the accumulation factor process exceeds $\rho$ within the day. In this case probabilities (\ref{fptcaso2}), (\ref{fptnk}) and (\ref{fptnk-}) will be equal to zero. Consequently, the fpt distribution will have non zero values only for $1\leq t \leq n-1$. In this case (\ref{fptI}) satisfies the following simpler equation:
\begin{equation}
\label{simple}
\begin{aligned}
& R_{i}(v,t;\rho)=1_{\{(1+i\Delta)^{t}<\rho\}}\bigg(\frac{1-H_{i}(t+v)}{1-H_{i}(v)}\bigg)+\\
& \sum_{a\in E}\sum_{m=1}^{t}\!\!\frac{b_{ia}(v+m)}{1-H_{i}(v)}1_{\{(1+i\Delta)^{m}<\rho\}}R_{a}(0,t-m;\frac{\rho}{(1+i\Delta)^{m}}). 
\end{aligned}
\end{equation}  
which is obtained from (\ref{fptI}) through relation (\ref{numerouno}).

\subsection{The intraday autocorrelation function}
In this subsection we derive the equation for the intraday autocorrelation function. Let us denote by  
\begin{equation}
\label{autocorr}
\begin{aligned}
& \gamma_{i}(x,v;t,s)=\\
& Cov(M_{x}(x+t+1),M_{x}(x+t+s+1)|Z(x)=i,B(x)=v).
\end{aligned}
\end{equation}
\indent From now on we will work under the assumption that $Kn\leq x\leq x+t\leq x+t+s < (K+1)n$. This means that the autocorrelation function is analyzed for times within the same day; for this reason we will refer to it as the intraday autocorrelation function.\\
\indent Notice that, because the semi-Markov process $Z(t)$ is time-homogeneous, the autocorrelation function (\ref{autocorr}) can be equivalently expressed independently of $x$ in the following simpler form:
\begin{equation}
\label{autocorrII}
\begin{aligned}
& \gamma_{i}(v;t,s)=\\
& Cov(M_{0}(t+1),M_{0}(t+s+1)|Z(0)=i,B(0)=v)=\\
& Cov_{(i,v)}(\prod_{k=0}^{t}(1+W(k)),\prod_{k=0}^{t+s}(1+W(k))).
\end{aligned}
\end{equation}
\indent To compute the autocorrelation function (\ref{autocorrII}) we need the knowledge of the expected accumulation factor denoted by
\begin{equation}
\label{momI}
m_{i}(v;t)=E[M_{0}(t+1)|(i,v)].
\end{equation}
Since $1\leq t \leq n-1$ we have that $W(k)=Z(k)$ and then 
\begin{displaymath}
m_{i}(v;t)=E[\prod_{k=0}^{t}(1+Z(k))|(i,v)].
\end{displaymath}
\indent Let us consider the random variable $\prod_{k=0}^{t}(1+Z(k))$; it is possible to give a recursive representation of this random variable. In fact 
\begin{equation}
\label{rec}
\begin{aligned}
& \prod_{k=0}^{t}(1+Z(k)) \overset{d}{=} 1_{\{T_{1}^{z}>t|Z(0)=i,B(0)=v\}}\prod_{k=0}^{t}(1+i\Delta)+\\
& \sum_{a\in E}\sum_{\theta =1}^{t}\prod_{k=0}^{t}(1+Z(k))1_{\{T_{1}^{z}=\theta, J_{1}^{z}=a|Z(0)=i,B(0)=v\}}    
\end{aligned}
\end{equation}
\noindent where the simbol $A\overset{d}{=}B$ denotes that the two random variables $A$ and $B$ have the same distribution.\\ 
\indent By taking the expectation in (\ref{rec}) and by using the independence between $1_{\{T_{1}^{z}=\theta, J_{1}^{z}=a|Z(0)=i,B(0)=v\}}$ and $\prod_{k=\theta}^{t}(1+Z(k))$ given the information set $\{Z(\theta)=a, B(\theta)=0\}$, we get
\begin{displaymath}
\begin{aligned}
& m_{i}(v;t)=(1+i\Delta)^{t+1}P(T_{1}^{z}>t|(i,v))+\\
& \sum_{a\in E}\sum_{\theta =1}^{t}(1+i\Delta)^{\theta}
P(T_{1}^{z}=\theta, J_{1}^{z}=a|(i,v))\\
& E[\prod_{k=\theta}^{t}(1+Z(k))|Z(\theta)=a,B(\theta)=0]    
\end{aligned}
\end{displaymath}
that is 
\begin{equation}
\begin{aligned}
\label{momI}
& m_{i}(v;t)=(1+i\Delta)^{t+1}\frac{1-H_{i}(t-1+v)}{1-H_{i}(v)}+\\
& \sum_{a\in E}\sum_{\theta =1}^{t}(1+i\Delta)^{\theta}\frac{b_{ia}(\theta+v)}{1-H_{i}(v)}m_{a}(0;t-\theta).    
\end{aligned}
\end{equation}
To evaluate the autocorrelation function we need also the knowledge of the second order cross moment of the accumulation factor
\begin{equation}
\label{momII}
m_{i}^{(2)}(v;t,s)=E_{(i,v)}[\prod_{k=0}^{t}(1+Z(k))\prod_{k=0}^{t+s}(1+Z(k))].
\end{equation}
Also in this case we can give a recursive representation of the random variable $\prod_{k=0}^{t}(1+Z(k))\prod_{k=0}^{t+s}(1+Z(k))$. In fact it holds true that    
\begin{equation}
\label{recII}
\begin{aligned}
& \prod_{k=0}^{t}(1+Z(k))^{2}\prod_{k=t+1}^{t+s}(1+Z(k))\overset{d}{=}\\
& 1_{\{T_{1}^{z}>t|(i,v)\}}(1+i\Delta)^{2(t+1)}(1+i\Delta)^{t-s}+\\
& \sum_{a\in E}\sum_{\theta =t+1}^{t+s}1_{\{T_{1}^{z}=\theta, J_{1}^{z}=a|(i,v)\}}\cdot \\
& (1+i\Delta)^{2(t+1)}(1+i\Delta)^{\theta-t}\prod_{k=\theta+1}^{t+s}(1+Z(k))+\\    
& \sum_{a\in E}\sum_{\theta =1}^{t}1_{\{T_{1}^{z}=\theta, J_{1}^{z}=a|(i,v)\}}\cdot \\       & (1+i\Delta)^{2(\theta+1)}\prod_{k=\theta+1}^{t}(1+Z(k))^{2}\prod_{k=t+1}^{t+s}(1+Z(k))
\end{aligned}
\end{equation}
\indent By taking the expectation in (\ref{recII}) and by using the independence between $1_{\{T_{1}^{z}=\theta, J_{1}^{z}=a|(i,v)\}}$ and the random variables $[\prod_{k=\theta+1}^{t+s}(1+Z(k))]$ and $[\prod_{k=\theta+1}^{t}(1+Z(k))^{2}\prod_{k=t+1}^{t+s}(1+Z(k))]$ given the information set $\{Z(\theta)=a, B(\theta)=0\}$, we get the following recursive equation for the second order cross moment
\begin{equation}
\label{momIIeqn}
\begin{aligned}
& m_{i}^{(2)}(v;t,s)=\frac{1-H_{i}(t+s+v)}{1-H_{i}(v)}(1+i\Delta)^{3t+2-s}+\\
& \sum_{a\in E}\sum_{\theta =t+1}^{t+s}\frac{b_{ia}(\theta+v)}{1-H_{i}(v)}(1+i\Delta)^{t+\theta+2}m_{a}(0;t+s-\theta)+\\    
& \sum_{a\in E}\sum_{\theta =1}^{t}\frac{b_{ia}(\theta+v)}{1-H_{i}(v)}(1+i\Delta)^{2(\theta+1)}m_{a}^{(2)}(0;t-\theta,s).
\end{aligned}
\end{equation}
By solving equations (\ref{momI}) and (\ref{momIIeqn}) we can obtain the intraday autocorrelation volatility function through the following relation:
\begin{equation}
\label{autofinale}
\gamma_{i}(v;t,s)=m_{i}^{(2)}(v;t,s)-m_{i}(v;t)m_{i}(v;t+s).
\end{equation}

\section{Application to real high frequency data}

\subsection{Database description}
The data we used in this work are tick-by-tick quotes of indexes and stocks downloaded 
from $www.borsaitaliana.it$ for the period January 2007-December 2010 (4 full years). 
The data have been re-sampled to have 1 minute frequency. Consider a single day (say day $k$ with $1 \le k \le d$)  
where $d$ is number of traded days in the time series. In our case we consider 
four years of trading (from the first of 
January 2007 corresponding to $d=1076$).
The market in Italy fixes the opening price at a
random time in the first minute after 9 am,
continuous trading starts immediately after and ends just before  5.25 pm,
finally the closing price is fixed just after 5.30 pm.
Therefore, let us define $S(t)$ as the price of the last trading
before 9.01.00 am , $S(t+1)$ as the price of the last trading
before 9.02.00 am and so on until
$S(nk)$ as the price of the last trading
before 5.25.00  pm.  
If there are no transactions in the minute,
the price remains unchanged
(even in the case the title is suspended and reopened
in the same day).
Also define
$S(nk+1)$ as the opening price and $S(nk)$ as the closing price.
With this choice $n=507$.
There was a small difference before the 28th of September 2009 since
continuous trading started at 9,05 am, and therefore
prior of that date we have $n=502$.
Finally, if the title has a delay in opening
or it closes in advance (suspended but not reopened),
only the effective trading minutes 
are taken into account. In this case $n$ will be smaller then 507.
The number of returns analyzed is then roughly 508000 for each stock.
We analyzed all the stocks in the FTSEMIB which are the 40 most capitalized
stocks in the Italian stock market.\\

\begin{figure}
\centering
\includegraphics[width=8cm]{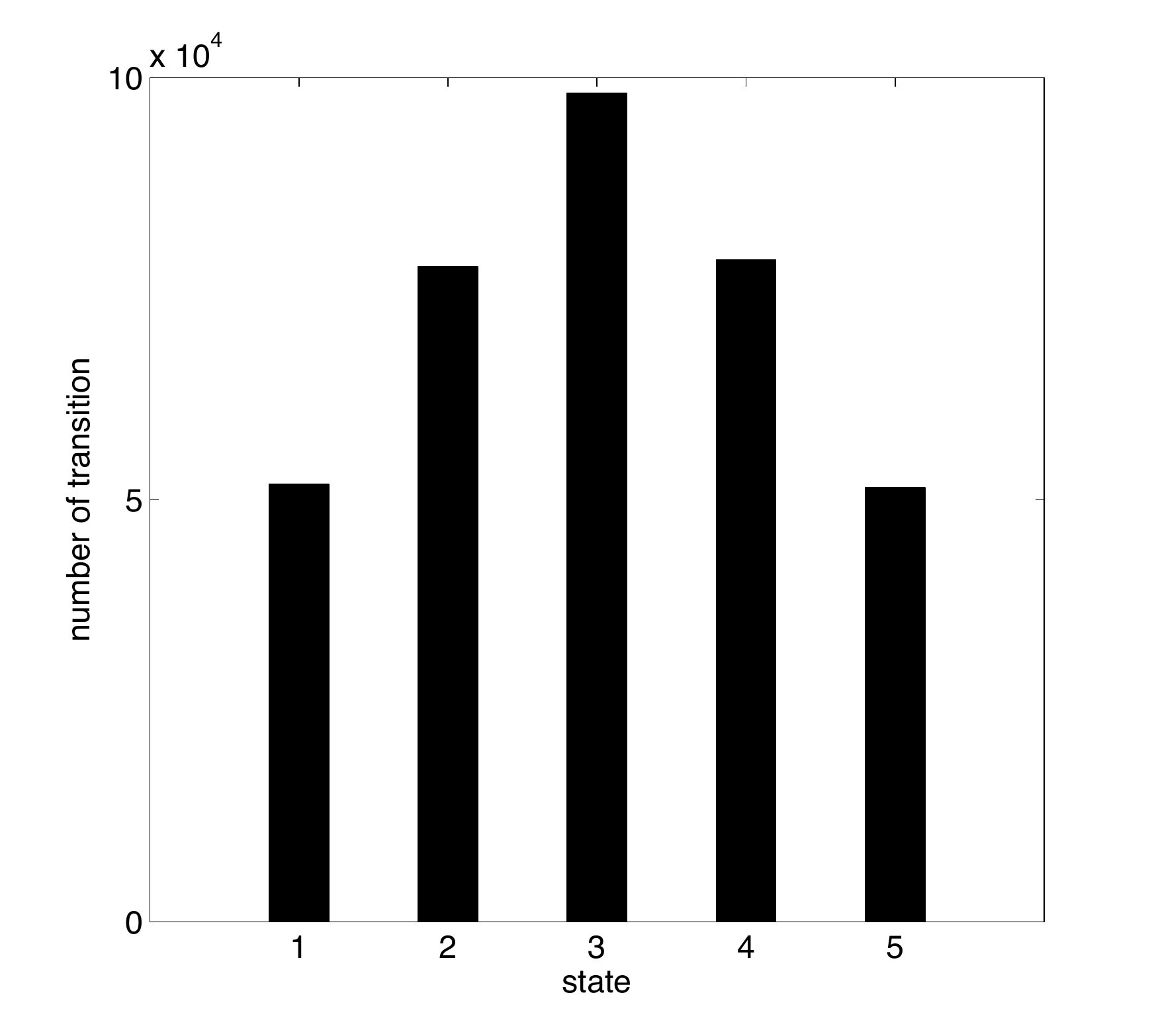}
\caption{Number of transition for the embedded Markov chain} \label{tra}
\end{figure}

To be able to model returns as a semi-Markov process the state space has to be discretized.
In the example shown in this work we discretized returns into 5 states chosen to be symmetrical with respect to returns equal zero. Returns are in fact already discretized in real data due to the discretization of stock prices. We then tried to remain as much as possible close to this discretization. In Figure \ref{tra} we show an example of the number of transition from state $i$
to all other states for the embedded Markov chain.

From the discretized returns we estimated the probabilities ${\bf P}$ and $G_{ij}(t)$ to generate a synthetic trajectory of the semi-Markov process modeled as described in Section \ref{price}. For comparison reason, we also generated a synthetic trajectory which follows a simple Markov model with transition probability matrix estimated from the real data. We then ended up with three trajectory: one representing real data, the second one a semi-Markov trajectory and the last one a Markov chain.
The three time series are used in the following to compare results on fpt distributions and on autocorrelations.

\subsection{Test}

The semi-Markov hypothesis is tested applying a test of hypothesis proposed by \cite{ste06} and shortly described here below. As already stated, the model can be considered semi-Markovian if the sojourn times are not geometrically distributed. The probability distribution function of the sojourn time in state $i$ before making a transition in state $j$ has been denoted by $G_{ij}(\cdot)$. Define the corresponding probability mass function by 
\begin{eqnarray}
\label{pmf}
&&g_{ij}(t)=P\{T_{n+1}-T_{n}= t|J_{n}=i, J_{n+1}=j\}=\nonumber \\ 
&&\left\{
                \begin{array}{cl}
                       \ G_{ij}(t)-G_{ij}(t-1)  &\mbox{if $t > 1$}\\
                         G_{ij}(1)  &\mbox{if $t=1$}\\
                   \end{array}
             \right.
\end{eqnarray}
Under the geometrical hypothesis the equality $g_{ij}(1)(1-g_{ij}(1))-g_{ij}(2)=0$ must hold, then a sufficiently strong deviation from this equality has to be interpreted as an evidence in favor of the semi-Markov model. The test-statistic is the following:
\begin{equation}
\label{test}
\hat{S}_{ij}=\frac{\sqrt{N(i,j)}\big(\hat{g}_{ij}(1)(1-\hat{g}_{ij}(1))-\hat{g}_{ij}(2)\big)}{\sqrt{\hat{g}_{ij}(1)(1-\hat{g}_{ij}(1))^{2}(2-\hat{g}_{ij}(1))}}.
\end{equation}
\noindent where $N(i,j)$ denotes the number of transitions from state $i$ to state $j$ observed in the sample and $\hat{g}_{ij}(x)$ is the empirical estimator of the probability $g_{ij}(x)$ which is given by the ratio between the number of transition from $i$ to $j$ occurring exactly after $x$ unit of time and $N(i,j)$. This statistic, under the geometrical hypothesis $H_{0}$ (or markovian hypothesis), has approximately the standard normal distribution, see \cite{ste06}.\\
\indent We applied this procedure to our data to execute tests at a significance level of $95\%$. Because we have $5$ states we estimated the  $5\times (5-1)$ waiting time distribution functions and for each of them we computed the value of the test-statistic (\ref{test}). The geometric hypothesis is rejected for $15$ of the $20$ distributions. Due to lack of space, we do not report all the values of the test-statistic, but they
are available upon request. In Table 1 we show the results of the test applied to the waiting time distribution functions with starting state $i=3$.\\
\begin{table}
\label{table1}  
\begin{center}
\begin{tabular}{llll}
\hline\noalign{\smallskip}
\textbf{state} & \textbf{state} & \textbf{score} & \textbf{decision} \\
\noalign{\smallskip}\hline\noalign{\smallskip}
$i=3$ & $j=1$ & 9,638 & $H_{0}$ rejected \\
$i=3$ & $j=2$ & 13,752 & $H_{0}$ rejected \\
$i=3$ & $j=4$ & 13,527 & $H_{0}$ rejected \\
$i=3$ & $j=5$ & 10,199 & $H_{0}$ rejected \\
\noalign{\smallskip}\hline
\end{tabular}
\caption{Results of the Test}
\end{center}
\end{table}
\indent The large values of the test statistic suggest the rejection of the Markovian hypothesis in favor of the more general semi-Markov one.\\

\subsection{Results on first passage time distribution}

For each of the stocks in our database we estimate the first passage time
distribution directly from the data (real data) and from the two synthetic time series generated as described above.\\
It is not possible to show all the results here, we then show only one figure of the fpt distribution
obtained for one stock (FIAT) and one value of $\rho$ (1.005). 

\begin{figure}
\centering
\includegraphics[width=8cm]{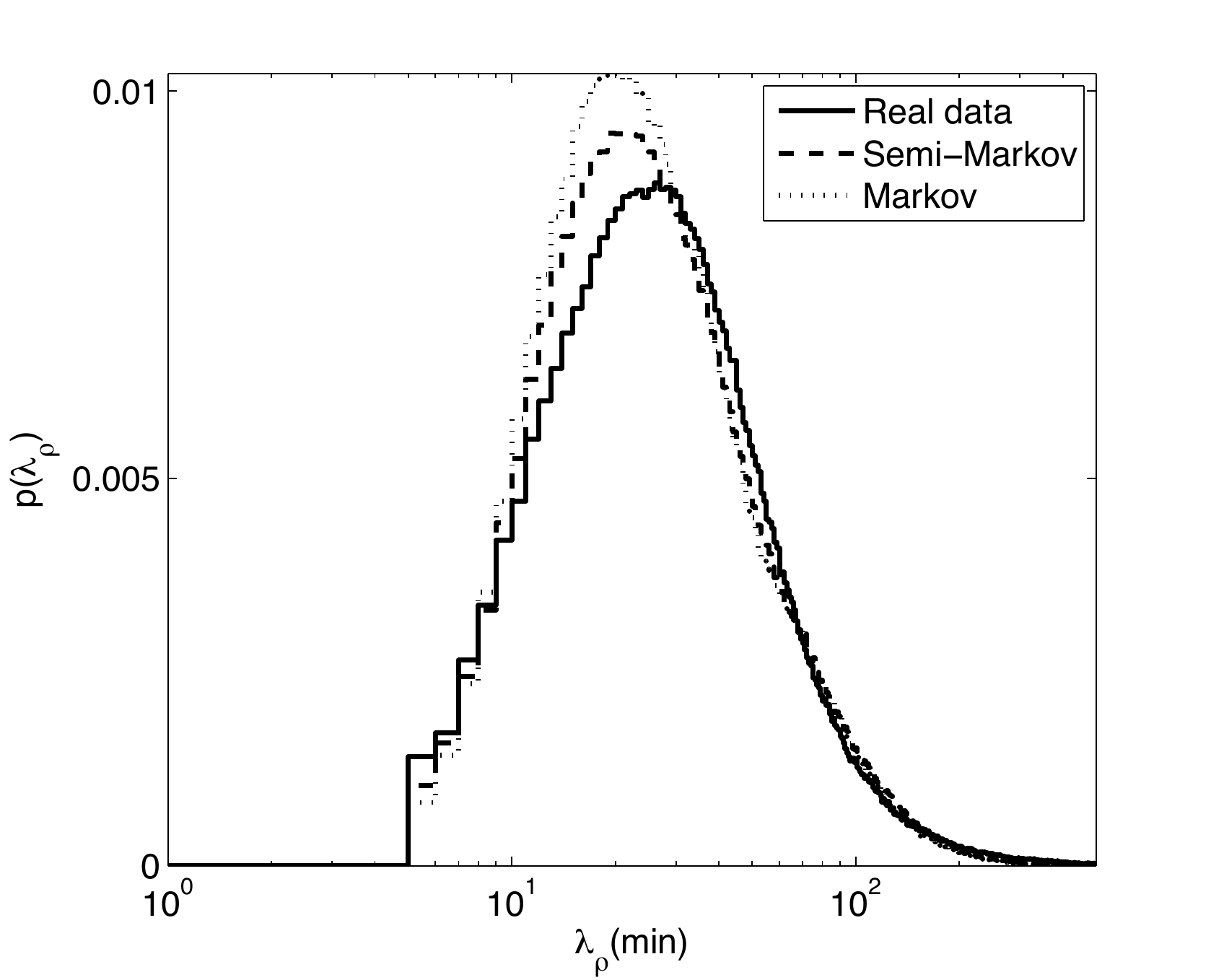}
\caption{First passage time distribution for $\rho=1.005$} \label{fpt}
\end{figure}

From Figure \ref{fpt} it is obvious that even if the semi-Markov model does not resemble exactly the fpt distribution of real data it works much better than the simpler Markov model. 

\subsection{Results on autocorrelation function}
\begin{figure}
\centering
\includegraphics[width=8cm]{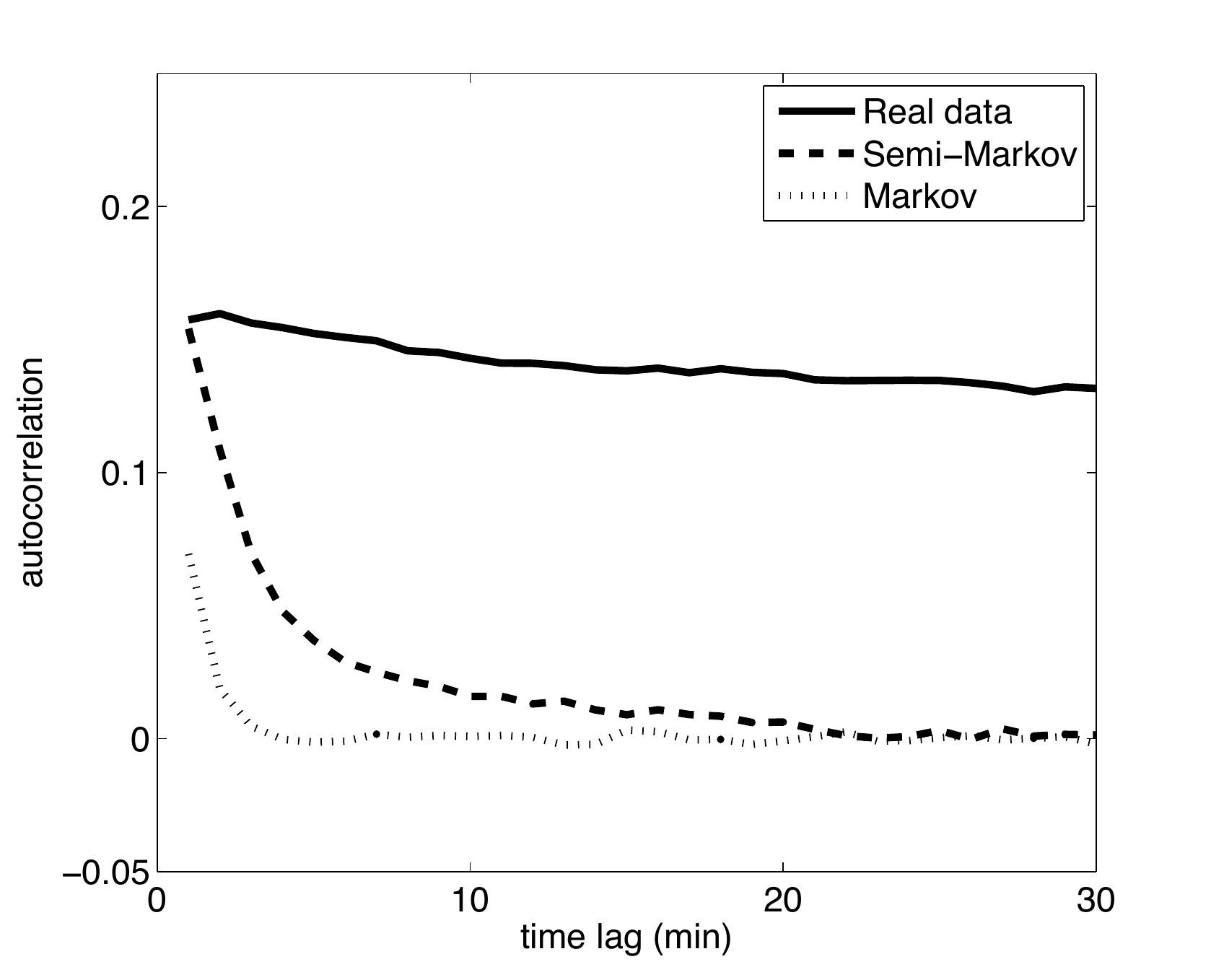}
\caption{Autocorrelation function of $W^2(t)$}\label{auto}
\end{figure}

Another important feature of the stochastic process that describes financial time series is
the autocorrelation of the square of returns. Indeed, while returns are uncorrelated the
absolute value or, which is the same, their square value is autocorrelated with a specific
decaying structure. This is well observed in our data as shown, again just for one stock, in Figure \ref{auto}.
In the figure we also compare the results obtained directly from data and those obtained by the use of our semi-Markov model.\\
\indent The autocorrelation of the square of returns is defined as 
\begin{equation}
\label{autosquare}
\Sigma(t, t+\tau)=Cov(W^2(t+\tau),W^2(t))
\end{equation}
To compute $(\ref{autosquare})$ observe that:
\begin{equation}
\label{prodmom}
\begin{aligned}
& E_{(i,v)}[W^2(t+\tau)W^2(t)]\\
& =\sum_{j\in E}\sum_{h\in E}\sum_{v'=0}^{t+v}\,^{b}\phi_{ij}^{b}(v;v',t)\,^{b}\phi_{hj}(v';\tau)j^{2}h^{2},
\end{aligned}
\end{equation}
\begin{equation}
\label{momuno}
E_{(i,v)}[W^2(t+\tau)]=\sum_{j\in E}\,^{b}\phi_{ij}(v;t+\tau)j^{2},
\end{equation}
\begin{equation}
\label{momdue}
E_{(i,v)}[W^2(t)]=\sum_{h\in E}\,^{b}\phi_{ih}(v;t)h^{2}.
\end{equation}
\indent If we assume that the process is in the stationary regime then $\Sigma(\tau):=\lim_{t \rightarrow \infty}\Sigma(t, t+\tau)$ is independent of $t$ and can be expressed by using the stationary distribution of the Markov chain $(W(t),B(t))$ studied in \cite{cri08}. The following formulas allows the computation of the autocovariance:
\begin{equation}
\label{prodmomII}
\begin{aligned}
& E_{(i,v)}[W^2(t+\tau)W^2(t)]\\
& =\sum_{j\in E}\sum_{h\in E}\sum_{v'\geq 0}\,\pi_{h}(v') ^{b}\phi_{hj}(v';\tau)j^{2}h^{2},
\end{aligned}
\end{equation}
\begin{equation}
\label{momunoII}
E_{(i,v)}[W^2(t+\tau)]=\sum_{j\in E}\,\pi_{j}j^{2},
\end{equation}
\begin{equation}
\label{momdueII}
E_{(i,v)}[W^2(t)]=\sum_{h\in E}\,\pi_{h}h^{2},
\end{equation}
\noindent where $\pi_{h}(v')=\frac{1-H_{i}(v')}{\mu_{ii}}$, $\pi_{j}=\sum_{v'\geq 0}\pi_{j}(v')$ and $\mu_{ii}$ is the mean recurrence time of state $i$ for the semi-Markov process $W(t)$.

Again we show in the figure that the semi-Markov model, even if is still far from the results on real data, can give much better results than a simple Markov-model.  

\section{Conclusions}
In this work we introduced a semi-Markov process to model high frequency stock returns. The model has been used to obtain both theoretical and empirical results on the first passage time distribution and on the autocorrelation function of the square of returns. We were able to calculate analytically the fpt distribution and the autocorrelation function and also to generate a synthetic time series starting from real data. We have shown, by means of Montecarlo simulations, that the semi-Markov model is able to reproduce much better than a simple Markov model results seen on real data. This suggest that the semi-Markov environment should be preferred when modeling stock market.


\begin{thebibliography}{99}

\bibitem{mai00} F. Mainardi, M. Raberto, R. Gorenflo, E. Scalas. Physica A 287 (2000) 468.
\bibitem{rab02} M. Raberto, E. Scalas, F. Mainardi. Physica A 314 (2002) 749.
\bibitem{dam05} G. D'Amico, J. Janssen, R. Manca. Decis. Econ. Finance 28 (2005) 79.
\bibitem{dam09} G. D'Amico, J. Janssen, R. Manca. Physica A 388 (2009) 3181.
\bibitem{sil04} D.S. Silvestrov, F. Stenberg. Commun. Stat. Theory Method 33 (2004) 591.
\bibitem{gui97}D.M. Guillaume, M.M. Dacorogna, R.R. DavVe, J.A. MWuller, R.B. Olsen, O.V. Pictet, Finance Stochast.1 (1997) 95
\bibitem{pet03} F. Petroni, M. Serva, Eur. Phys. J. B 34 (2003) 495.
\bibitem{ste06} F. Stenberg, R. Manca, D.S. Silvestrov. Theory Stoch. Processes 12(28) no 3-4 (2006).
\bibitem{sim02} I. Simonsen. M.H. Jensen, A. Johansen, Eur. Phys. J. 27 (2002) 583.
\bibitem{jen04} M. H. Jensen, A. Johansen, F. Petroni, I. Simonsen. Physica A 340, (2004) 678.
\bibitem{lim01} N.Limnios, G.Opri\c{c}an, Semi-Markov Processes and Reliability Modeling, Birkh\"{a}user, Boston. (2001).
\bibitem{jan06} J. Janssen, R. Manca. Applied semi-Markov Processes, Springer (2006).
\bibitem{bar04} V. Barbu, M. Boussemart, N. Limnios. Commun. Stat. Theory Method 33 (2004) 2833.
\bibitem{dam10} G. D'Amico, J. Janssen, R. Manca. Methodol Comput Appl Probab. 12(2) (2010) 215.
\bibitem{cri08} O. Chryssaphinou, M. Karaliopoulou, N. Limnios. Commun. Stat. Theory Method 37 (2008) 1306.
\end{thebibliography}
\end{document}